# Deep Learning Approaches for Extracting Adverse Events and Indications of Dietary Supplements from Clinical Text

Yadan Fan[1], Sicheng Zhou[1], Yifan Li[2], Rui Zhang[1,2]

[1]Institute for Health Informatics, [2]College of Pharmacy, University of Minnesota, Minneapolis, MN, USA

* Corresponding author: Rui Zhang, PhD

Address: 8-100 PWB, 516 Delaware St SE, Minneapolis, MN 55446 USA

E-mail: zhan1386@umn.edu

Tel.: 612-626-8654

Email addresses:

YF: fanxx421@umn.edu

SZ: zhou1281@umn.edu

YL: li000426@umn.edu

RZ: zhan1386@umn.edu

# ABSTRACT

**Objective**

The objective of our work is to demonstrate the feasibility of utilizing deep learning models to extract safety signals related to the use of dietary supplements (DS) in clinical text.

**Methods**

Two tasks were performed in this study. For the named entity recognition (NER) task, Bi-LSTM-CRF (Bidirectional Long-Short-Term-Memory Conditional Random Fields) and BERT (Bidirectional Encoder Representations from Transformers) models were trained and compared with CRF model as a baseline to recognize the named entities of DS and Events from clinical notes. In the relation extraction (RE) task, two deep learning models, including attention-based Bi-LSTM and CNN (Convolutional Neural Network), and a random forest model were trained to extract the relations between DS and Events, which were categorized into three classes: positive (i.e., indication), negative (i.e., adverse events), and not related. The best performed NER and RE models were further applied on clinical notes mentioning 88 DS for discovering DS adverse events and indications, which were compared with a DS knowledge base.

**Results**

For the NER task, deep learning models achieved a better performance than CRF, with F1 scores above 0.860. The attention-based Bi-LSTM model performed the best in the relation extraction task, with the F1 score of 0.893. When comparing DS event pairs generated by the deep learning models with the knowledge base for DS and Event, we found both known and unknown pairs.

**Conclusion**

Deep learning models can detect adverse events and indication of DS in clinical notes, which hold great potential for monitoring the safety of DS use.

# INTRODUCTION

The popularity of dietary supplements (DS) has continued to grow during recent years. A 2019 survey conducted by the Council for Responsible Nutrition indicates that the use of DS remains strong and increasing, with 77% of Americans taking DS, up from 66% compared with 2008 [1]. Despite the widespread use and consumers' increasingly receptive attitudes, there still exist some quality, efficacy, and safety issues, such as insufficient information on the identity of ingredients, lack of well-designed human clinical trials to assess the safety of DS, limited in-vitro experiments to elucidate the mechanisms for actions, etc [2]. Due to the complex regulatory environment for DS in the United States, some DS ingredients may not have undergone thorough safety evaluations before being legally introduced into the market since DS are considered as a special category of food. However, adverse events (AEs) related to DS can be severe or even deadly. According to one study, among the total AEs (n = 15,430) submitted to the US Food and Drug Administration (FDA) Center for Food Safety and Applied Nutrition Adverse Event Reporting System (CAERS) during 2004-2013, 25.4% resulted in hospitalization and 2.2% led to deaths [3]. The lack of premarket safety requirements combined with the perception by the public that most DS products are natural and therefore safe have further contributed to the paucity of voluntary reporting data regarding AEs through post-market surveillance mechanisms. Moreover, the reporting data on AEs may suffer from the lack of accurate information on the temporal relationship between product ingestion and the onset of AEs. Such reporting bias makes detection of some potential causal relationships between DS and AEs difficult [4].

Due to the limitations mentioned above, there remains a critical need for use of alternative data sources for overseeing and monitoring safety in terms of DS use. It has been long recognized that EHR data, especially clinical notes, provide the most comprehensive documentation of clinical events that occur during the course of health care [5]. Compared with conventional data sources (i.e., clinical trials, spontaneous reporting data), clinical notes present several advantages such as the availability of more comprehensive, real-world patient information and accurate documentation of disease development. Clinical Natural Language Processing (NLP) techniques have been extensively leveraged to approach this task through performing extraction on medication entities and their relationships with corresponding AEs. Over the years, various clinical NLP shared tasks, such as the i2b2 (Informatics for Integrating Biology and the Bedside) challenge [6], n2c2 (National NLP Clinical Challenges) [7], and the most recent MADE (the 2018

Medication and Adverse Drug Event) challenge [8], have been organized to examine the state-of-the-art NLP methods for clinical concept recognition and relation extraction. The approaches for the medication entity extraction task, namely named entity recognition (NER) task, fall into categories of rule-based [9], supervised machine learning [10], and hybrid methods [11]. The most recent development in deep learning techniques have achieved more competitive results compared with traditional machine learning methods. Particularly, bidirectional long short-term memory (Bi-LSTM) models combined with a Conditional Random Fields (CRF) layer have been shown to achieve better performance in medical concept extraction [12, 13]. Existing methods for relation extraction (RE) can be grouped into rule-based, bootstrapping, supervised, distant supervision, unsupervised, and deep learning methods. The methods in the last decade have been dominated by feature-based and kernel-based methods [14], where the hand-designed linguistic features were fed into machine learning classifiers such as logistic regression classifiers and support vector machines. However, compared with state-of-the-art deep learning methods, supervised machine learning techniques rely heavily on hand-crafted features and language specific resources, which are more time consuming and labor-intensive to construct.

Using DS could lead to various AEs caused by individual DS use or their interactions with other concomitant DS, drugs and food due to their complicated characteristics [15]. Several studies have developed methods for the creation of DS terminology and knowledge base as well as the detection of DS associated AEs from different data sources [17-23]. For example, we have extracted and standardized DS information from online sources to build an integrated DS knowledge base, (i.e., iDISK) [16]. And we demonstrated that as compared with the UMLS, DS terminology in the iDISK contains more novel synonyms, and achieved a better performance in a DS NER task on biomedical literature [17]. Also, we demonstrated the utility of word embeddings on clinical notes for DS terminology expansion [18]. Previously, we employed signal detection methods to extract AEs associated with DS from Center for Food Safety and Applied Nutrition (CFSAN) AE reporting system (CAERS) [19]. We developed methods to extract the DS usage information from Twitter, and assessed the association between DS use and mental disorders (e.g., anxiety, depression) [20]. We also mined AEs from DS product labels in the Dietary Supplement Label Database (DSLD) using topic modeling [21]. In addition, we developed a rule-based NLP system to normalize DS product names in DSLD [22]. Another study successfully applied a deep neural network to identify the drug-drug interactions and drug-food interactions based on their structural information and names [23]. However, no studies have investigated the detection of DS and related AEs in clinical notes. Like medications, a great deal of DS information including their associated indications and AEs is documented in clinical notes. Recognizing DS named entities and their relationships with signs and symptoms in clinical notes is of great significance for automatic safety surveillance on DS. Detecting AEs related to DS use is critical for patient safety. Thus, applying NLP techniques for automatic AEs extraction can accelerate downstream pharmacovigilance-related research.

Thus the main contributions of this study are:
- To the best of our knowledge, this is the first study of deep learning models for extracting AEs and indications associated with DS from clinical notes
- An evaluation of different deep learning (e.g., pre-trained BERT) models on annotated DS-specific clinical corpora
- Demonstration of the feasibility of deep learning models applied to clinical notes to facilitate discovery of DS safety knowledge

## METHODS

This study consists of two tasks: NER and RE. The methods for these two tasks are described in detail as follows.

### Task 1- NER

**Study design**

The NER task was carried out in the following five steps: (1) preprocessing the clinical notes and randomly collecting 1000 sentences for each of the 7 commonly used DS; (2) annotating collected sentences with mentions of DS, indications, and AEs to generate the gold standard; (3) randomly splitting the gold standard into training, development, and test sets; (4) training, tuning and evaluating the models; (5) comparing the performance of deep learning models and a baseline CRF model.

**Data collection and annotation**

The dataset used in this study was collected from clinical data repository (CDR) of the academic medical center affiliated with the University of Minnesota (UMN). The CDR contains 180 million clinical notes of over 2.9 million patients seeking health care at 8 hospitals and over 40 clinics. IRB (institutional review board) was obtained for accessing the clinical notes. Document-level clinical notes were split into sentences through sentence boundary detection using BioMedICUS (BioMedical Information Collection and Understanding System), an NLP pipeline developed at UMN [24]. Based on our prior study on DS term expansion [18], a collection of DS terms were used for retrieving sentences mentioning 7 DS, including black cohosh, chamomile, cranberry, folic acid, garlic, turmeric, and valerian, which were chosen based on their popularity in the CDR. In total, 7,000 sentences (1,000 sentences for each of the 7 DS) were randomly selected from the resulting sentence-level corpus. Annotation guidelines were created based on 100 randomly selected sentences out of the sentence-level corpus of 7,000 sentences for this NER task. Disagreement was resolved with discussion to reach consensus. The inter-rater agreement was calculated based on these 100 sentences using Cohen's kappa score. The remaining sentences were equally split into two parts, which were independently annotated by two experts with clinical background (each annotated 3,450 sentences). "Beginning-Inside-Outside" (BIO) annotation schema was used. Among these 7,000 sentences, 2,812 (40.17%) have both mentions of DS and Events. Two categories of named entities were defined and annotated: DS and Event. DS is defined as any mention of DS, including generic (i.e., black cohosh) and brand names (i.e., garlique), synonyms (i.e., folate), abbreviations (i.e., cran) and misspellings (i.e., tumeric). Event includes indications (a sign or symptom for which DS is taken for, for example, black cohosh for hot flash), AEs (a sign or symptom caused by DS use, for example, liver damage caused by black cohosh), and other signs or symptoms (not related to the DS use).

**Models**

The standard neural model for the NER task is based on Bi-LSTM-CRF. In this model, word or character embeddings or the combination of them are often used as inputs. In this task, we compared three BiLSTM-CRF [25] models using three types of inputs (word embeddings only, word embeddings combined with CNN character-level representations, word embeddings combined with LSTM character-level representations) with CRF model as a baseline to extract named entities of DS and Events from clinical narratives. The BiLSTM-CRF model consists of three layers, including an embedding layer, a bidirectional LSTM layer and a CRF layer. The sequence of embeddings is given as the input for the bi-LSTM layer, which then returns a representation of the left and the right context for each word. These representations are further concatenated and linearly projected onto a CRF layer. In this study, word-level representations are distributed word embeddings trained from a large clinical corpus using word2vec in one of our previous studies [18]. Specifically, these embeddings were trained using over 26 million clinical notes. Besides word-level information, character-level information was also considered since character-level embeddings have found to be beneficial for out-of-the-vocabulary words and are capable of capturing morphological information [26]. Both CNN [26] and RNN (Bi-LSTM) [25] models were applied to generate character-level representations separately. To be specific, character embeddings are randomly initialized for every character. The character embeddings corresponding to every character contained in a word are given as the input to a CNN or a Bi-LSTM model, the outputs of which are the character-level representations for each word. Character-level representations are further concatenated with word2vec word embeddings to be fed into the Bi-LSTM layer.

Besides the standard neural models, the NER task can also be approached using transfer learning. Transfer learning is the process of training a model on a large-scale dataset and then using the pre-trained model to conduct the learning on a task specific dataset, which might be smaller. The benefit of such method is that not much data is needed in the downstream task to achieve good results. BERT, which stands for Bidirectional Encoder Representations from Transformers, is one of the state-of-the art and empirically powerful language models released by Google in 2019 [27]. The key innovation of BERT [28] is to apply bidirectional training of transformers to language modeling. To train the language model, BERT utilizes two training strategies: Masked Language Model (MLM) and Next Sentence Prediction (NSP). The language models trained in such a way often have a deeper sense of language context, which can be further applied to handle a variety of NLP tasks (i.e., NER) with just one additional output layer. Utilizing the self-attention mechanism of the transformer encoder, BERT is shown to be able to capture syntactic and coreference information [29]. BERT also utilizes positional embeddings to incorporate sequential information in order to overcome the limitation imposed by self-attention. Additionally, the use of WordPiece embedding can help achieve a balance between size of the vocabulary and out-of-vocabulary tokens. In this study, we applied two pre-trained BERT models, Bert large cased model and Clinical BERT model to perform the NER task. Specifically, the pretrained BERT large cased model (i.e., 24 layers, 1024 hidden units, 16 attention heads, 340M parameters) was trained on the BooksCorpus (800M words) and English Wikipedia (2,500M words). Clinical BERT [30], initialized from BERT-base model (i.e., 12 layers, 768 hidden units, 12 attention heads, 110M parameters), was trained on MIMIC-III clinical notes.

**Model training and evaluation**

A total of 7,000 sentences were split into training (80%), development (10%), and test (10%) sets. We trained the deep learning models on the training set and applied it on the development set. The model with the best performance on the development set was further applied on the test set for final evaluation. We denote the Bi-LSTM-CRF model using only word embeddings for inputs as Bi-LSTM-CRF (word only). Different numbers of hidden units in the Bi-LSTM layer were tested (i.e., 64, 128, 256). The optimal hidden size was set as 256. We denote the Bi-LSTM-CRF model using CNN to generate character-level representations as Bi-LSTM-CRF (char cnn). We experimented with a range of hyperparameters, including the number of filters (i.e., 50, 100, 200, 300), the kernel size (i.e., 2, 3, 4, 5, 6), the hidden size for the LSTM layer (i.e., 32, 64, 128, 256). The optimal hyperparameters for this model are: kernel size of 5, filter numbers of 300, and 256 for the LSTM hidden size. The Bi-LSTM-CRF model using bi-LSTM to generate character-level information is denoted as Bi-LSTM-CRF (char lstm), which was experimented with various hyperparameters, including the hidden size of character Bi-LSTM layer (i.e., 32, 64), the hidden size of word Bi-LSTM layer (i.e., 32, 64, 128, 256). The Bi-LSTM hidden size for the character and token level were set as 25 and 256, respectively. Early stopping was used to reduce overfitting. To be specific, if the F1 score didn't increase within 500 training steps, the training process stopped. Furthermore, two pretrained BERT models were fine-tuned and evaluated using the training and development data, respectively. Specifically, the two BERT models were fine-tuned on training data for 3 epochs, with a learning rate of $2 \times 10^{-5}$. A CRF layer was built on the top of the BERT model to perform the NER task. The final performance was reported using the test data. We also trained a CRF model to compare against with the deep learning models. Features used to train the CRF model include word suffix, POS tags, the POS tags of the nearby words (one word before and one word after), etc. Precision, recall, and F1 score were used as the evaluation metrics.

**Task 2 – RE**

**Study design**

The relation extraction task was performed in the following steps: (1) randomly collecting 3,000 sentences on 15 DS; (2) annotating and categorizing DS and Event mention pairs into one of the three relations (i.e., positive, negative, and not related); (3) splitting the data into training, development, and test sets; (4) training, tuning, and evaluating models; (5) comparing the performances of deep learning and random forests; (6) applying the model with best performance on 88 unseen DS for knowledge discovery.

**Data collection and annotation**

In order to collect a corpus for the RE task which requires the co-occurrent of DS and Events mentions, a list of DS terms (similar to the NER task) and signs/symptoms terms compiled from iDISK [16] were used to randomly retrieve sentences. Based on the popularity and availability of DS in our CDR, we retrieved a total of 3,000 sentences (200 sentences on each) of the 15 DS, including black cohosh, chamomile, cranberry, dandelion, folic acid, garlic, ginger, ginkgo, ginseng, glucosamine, green tea, lavender, melatonin, milk thistle, and saw palmetto). We followed the annotation guideline of the NER task for annotating DS and Event entities. Three relation types were further defined between DS and Event entity pairs: positive, negative and not related. Positive means that a DS is taken for some Events (indications). Negative refers that the DS has caused some Events (AEs or side effects). The relation type "not related" indicates that there were no direct relationships between the DS and Event based on the semantic and linguistic cues given by the context in the sentence. Negation was considered when domain experts completed RE annotations. If the relationship between DS and Events was negated, we annotated it as "not related". However, we did not include the probabilistic terms in our annotation. 100 sentences were randomly selected and annotated by two annotators to evaluate the inter-rater agreement using Cohen's kappa score. The remaining sentences (2,900) were equally split and independently annotated by the two annotators.

**Models**

In this study, we compared two deep learning models with random forest as a baseline for relation extraction, including a CNN model and an attention-based Bi-LSTM model. The CNN model [31] consists of 4 layers: the embedding layer, the convolutional layer, the pooling layer, and the fully connected layer with softmax function to perform the final classification. For each word in the sentence, its word embedding was obtained through training a word2vec model on a large medical corpus in our previous study [18], was concatenated with two position embeddings, which encode information on the relative distance of the current word to the two entities of interest in the sentence. The dimensionality of the position embedding is a hyperparameter, which needs to be tuned. The convolution layer with varied filter sizes is applied to recognize n-gram features. The max pooling layer is further used to extract the most important or relevant features generated from the convolution layer. The max pooling scores from each filter were concatenated to form a single vector, which goes through a dropout and is fed into a fully connected layer.

The attention-based Bi-LSTM (Att-BLSTM) [32] model for relation extraction consists of 4 layers: the embedding layer with each word in a sentence representing by a pretrained word2vec word embedding from a previous study [18], the Bi-LSTM layer with the forward and backward LSTM outputs concatenating through element-wise sum, the attention layer which produces a weight vector to be multiplied with Bi-LSTM outputs, and the final output layer using softmax function. Dropout and L2 regularization are applied in the final output layer for reducing overfitting. Dropout is also applied in the embedding layer and LSTM layer for regularization. Specifically, the attention layer produces an attention vector which is equal to the length of the sequence. Each value in this vector is the weight associated with the corresponding Bi-LSTM output feature vector. The weighted linear combination of the Bi-LSTM outputs and attention weights form the output of the attention layer. With the addition of the attention layer, the model is capable of capturing more significant semantic features with decisive effects on the classification results.

**Model training and evaluation**

The dataset was divided into training (80%), development (10%) and test (10%) sets. The development set was used for tuning hyperparameters. For the CNN relation extraction model, the inputs for the model are the concatenation of the word embeddings for the current token and two position embeddings, one is the relative distance from the current token to the DS entity head and the other is the relative distance to the Event entity head. The three vectors are concatenated and fed into the model as inputs. Since the positional information is encoded in inputs, different convolutional filters can be learned for the same n-gram if it occurs in a different position relevant to the entities of interest. We experimented with a set of parameters, including the dimensionality of the position embedding (i.e., 50, 100, 200), the number of filters (i.e., 64, 128, 256, 512), filter sizes (i.e., 2, 3, 4, 5, 6). The optimal hyperparameters are as follows: position embedding dimension of 100, filter sizes of [2, 3, 4], 128 filters for each size. The dropout rate is 0.3. For the Att-BLSTM model, we tuned the hyperparameter hidden size of the Bi-LSTM layer (i.e., 64, 128, 256, 512). The optimal value for the hidden size was chosen as 128. The model with optimal parameters was applied to the test set for final model evaluation. Early stopping was used to reduce overfitting. Additionally, a random forest model was trained as a baseline model. Some preprocessing was performed, including normalization and stop words removal. N-gram features were used for training the model. Precision, recall, and F-1 score were used as the evaluation metrics.

**Knowledge discovery**

To compare the results generated by our methods with existing DS safety knowledge, we further collected sentences containing another 88 DS terms (listed in Supplementary Table 1) based on the popularity and availability of DS in the CDR. Specifically, the sentences mentioning the 88 DS were collected from over 26 million clinical notes ranging from April 2015 to December 2016 at the University of Minnesota Medical Center. Our trained NER model was applied to detect the mentions of DS and Events. The sentences with both mention of DS and Events were further fed into the best performing RE model. In each sentence, a DS and Event entity pair was classified into one of the three categories based on the best RE model. We analyzed the results and limited the scope to positive and negative relations. Given the frequencies of DS and Event entity pairs in these two categories, entity pairs with number of their source sentences larger than ten were further compared with the knowledge in the existing database, Natural Medicines Comprehensive Database (NMCD). NMCD is managed by the Therapeutic Research Center, which provides 15 categories (e.g., scientific names, indications, safety, effectiveness) of information for each product. In addition, we conducted a manual review of 20 randomly selected high and low frequency pairs with their 100 source sentences to estimate the performance of our deep learning algorithms.

## Results
### Data set

The Cohen's kappa scores for NER and RE are 0.879 and 0.863 respectively. There was a total of 12,213 DS and 3,807 Events entities in the 7,000 sentences of 7 DS. Among the 5,131 relation pairs, 3,451 are positive relations, 1,071 are negative relations, and 609 fell into the category of

"not related". Among the sentences with mentions of 88 DS, there were 31,675 of DS and Event entity pairs.

**Performance of NER models**

The results of NER models on the test set out of 7,000 sentences are shown in Table 1. According to the results, deep learning models outperformed the CRF model. Four deep learning models have very close F1 scores, although the BERT model performs slightly better overall. For DS entities, five models performed well, with F1 scores all over 0.8. However, the CRF model has a relatively low recall score in recognizing Event entities, partially due to the small number of Event entities used for training, which also indicates that the deep learning models are more resistant to imbalanced data. Compared with other models, the BERT model has significantly outperformed them on the NER task based on student t-test (P< 0.05).

Table 1. Results of NER models on the test set. We run all models 5 times and report mean ± standard deviation.

|  | DS | | | | Event | | | | Overall (micro) | | | |
| --- | --- | --- | --- | --- | --- | --- | --- | --- | --- | --- | --- | --- |
|  | P* | R* | F1 | Num* | P | R | F1 | Num | P | R | F1 | Num |
| CRF | 0.900 ± 0.00 | 0.791 ± 0.00 | 0.842 ± 0.00 | 1247 | 0.714 ± 0.00 | 0.567 ± 0.00 | 0.632 ± 0.00 | 356 | 0.861 ± 0.00 | 0.741 ± 0.00 | 0.797 ± 0.00 | 1603 |
| Bi-LSTM-CRF (word only) | 0.905 ± 0.002 | 0.854 ± 0.007 | 0.879 ± 0.003 | 1247 | 0.812 ± 0.015 | 0.825 ± 0.007 | 0.818 ± 0.009 | 356 | 0.884 ± 0.004 | 0.847 ± 0.003 | 0.865 ± 0.003 | 1603 |
| Bi-LSTM-CRF (char lstm) | 0.900 ± 0.006 | 0.860 ± 0.002 | 0.879 ± 0.003 | 1247 | 0.806 ± 0.008 | 0.837 ± 0.011 | 0.822 ± 0.008 | 356 | 0.877 ± 0.003 | 0.855 ± 0.003 | 0.866 ± 0.002 | 1603 |
| Bi-LSTM-CRF (char cnn) | 0.905 ± 0.006 | 0.864 ± 0.004 | 0.884 ± 0.003 | 1247 | 0.847 ± 0.018 | 0.845 ± 0.007 | 0.846 ± 0.011 | 356 | 0.892 ± 0.006 | 0.860 ± 0.003 | 0.876 ± 0.004 | 160 |
| Clinical BERT | 0.931 ± 0.002 | 0.845 ± 0.002 | 0.886 ± 0.002 | 1247 | 0.836 ± 0.014 | 0.840 ± 0.007 | 0.838 ± 0.008 | 356 | 0.908 ± 0.003 | 0.845 ± 0.002 | 0.875 ± 0.001 | 1603 |
| BERT | 0.931 ± 0.005 | 0.850 ± 0.003 | **0.889 ± 0.003** | 1247 | 0.860 ± 0.010 | 0.854 ± 0.006 | **0.857 ± 0.004** | 356 | 0.914 ± 0.007 | 0.851 ± 0.003 | **0.881 ± 0.003** | 1603 |

*P: Precision; R: Recall; Num: Number

**Performance of relation extraction models**

The results of relation extraction are shown in Table 2. Overall, two deep learning models achieved better performances than the random forest model. Specifically, the averaged F1 score of Att-BLSTM model (0.893) is higher than that of the CNN (0.890) model. However, based on student t-test (P<0.05), overall performances of the Att-BLSTM and CNN are not significantly different. Both methods significantly outperformed the Random Forest model.

Table 2. Results of relation extraction task on the test set. We run all models 5 times and report mean ± standard deviation.

| | Positive | | | | Negative | | | | Not related | | | | Overall (micro) | | | |
|---|---|---|---|---|---|---|---|---|---|---|---|---|---|---|---|---|
| | P | R | F1 | Num | P | R | F1 | Num | P | R | F1 | Num | P | R | F1 | Num |
| Random Forest | 0.835 ± 0.002 | 0.939 ± 0.003 | 0.884 ± 0.002 | 336 | 0.782 ± 0.009 | 0.716 ± 0.007 | 0.747 ± 0.006 | 109 | 0.825 ± 0.011 | 0.438 ± 0.006 | 0.572 ± 0.005 | 69 | 0.823 ± 0.003 | 0.824 ± 0.002 | 0.813 ± 0.002 | 514 |
| CNN | 0.937 ± 0.013 | 0.936 ± 0.031 | 0.936 ± 0.010 | 336 | 0.804 ± 0.057 | 0.926 ± 0.021 | 0.859 ± 0.026 | 109 | 0.824 ± 0.095 | 0.634 ± 0.060 | 0.721 ± 0.040 | 69 | 0.899 ± 0.013 | 0.896 ± 0.016 | 0.890 ± 0.016 | 514 |
| Att-BLSTM | 0.913 ± 0.011 | 0.967 ± 0.017 | **0.939 ± 0.004** | 336 | 0.869 ± 0.035 | 0.861 ± 0.063 | **0.863 ± 0.024** | 109 | 0.876 ± 0.028 | 0.798 ± 0.009 | **0.826 ± 0.007** | 69 | 0.897 ± 0.006 | 0.899 ± 0.005 | **0.893 ± 0.004** | 514 |

**Knowledge discovery**

Since the Att-BLSTM model has the best performance in relation extraction, it was further applied on the 13,474 sentences with mentions of 88 unseen DS extracted from more than 26 million clinical notes to categorize the DS and Event entity pairs into one of the three predefined classes. In total, there are 18,348 positive relations and 13,130 negative relations. We also checked the existence of these positive and negative DS-AE pairs with frequency larger than 10 by comparing with the information in the NMCD and indicated in the table. Within 133 positive signals, 94 (70.7%) are known in NMCD, and 39 (29.3%) are unknown signals. Among 84 negative signals, 48 (57.1%) and 36 (42.9%) are known and unknown signals in NMCD, respectively. Example unknown pairs are listed in the Supplementary Table 2. To further estimate the performance of our deep learning methods to detect signals, we randomly select 20 pairs (10 for positive and 10 for negative) and manually reviewed the randomly selected 10 sentences for each pair (200 total sentences). Details of these entity pairs with their frequency, precision, and example sentences are listed in the Table 3. At the supplement level, the precision for vitamin C, fish oil, vitamin E, peppermint, zinc, psyllium, biotin, and niacin are 90%, 70%, 100%, 95%, 100%, 70%, 100%, and 61.7%, respectively.

The findings generated by the deep learning models are consistent with the known knowledge regarding the indications or AEs of DS. For example, Vitamin C can promote wound healing because of its role in collagen formation. Vitamin C is a co-factor in proline and lysine hydroxylation, a necessary step in the formation of collagen. Rash, flushing, and hives are common side effects of niacin. The allergic symptoms to fish oil include rash, hives, diarrhea. Fish oil may inhibit platelet aggregation and may potentially increase the risk of bleeding. In addition, we found some unknown pairs which are worth further investigation.

Table 3. Selected DS and Event entity pairs from relation extraction on 88 DS with positive and negative categories. "√" indicating the existence in NMCD, "X" indicating not existed in NMCD. Precision is calculated based on the correctness of randomly selected 10 source sentences.

| Category | Positive (indications) | | | Negative (adverse events) | | |
|---|---|---|---|---|---|---|
| | Entity pair (frequency, precision) | NMCD | Example | Entity pair (frequency, precision) | NMCD | Example |
| Top 10 entity pairs | Peppermint, Nausea (203, 100%) | √ | • Patient has intermittent nausea which is relieved with peppermint and schedules Zofran.<br>• He has much less nausea with peppermint oil and marijuana. | Niacin, Rash (185, 90%) | √ | • Lisinopril causes a cough and niacin causes a rash.<br>• He lists his current allergies as a rash to niacin and swelling to penicillins. |
| | Fish oil, Hyperlipidemia (197, 80%) | √ | • Patient has history of hyperlipidemia which was until recently well-controlled with fish oil and simvastatin.<br>• I was told to resume fish oil for hyperlipidemia. | Niacin, Hives (117, 60%) | × | • Niacin causes hives and rash.<br>• Patient stating reaction to niacin is hives though has used mvi in past without issues. |
| | Vitamin C, Wound (194, 100%) | √ | • Consider ordering an additional 500 mg vitamin c daily and 10,000 iu vitamin A daily for wound healing support.<br>• Starting mv, Vitamin C and zinc for wound healing. | Fish oil, Bleeding (104, 90%) | √ | • Hold fish oil for hyperlipidemia due to risk of bleeding due to low platelets.<br>• He takes fish oil now and then and since this can increase bleeding he should hold this for until his colitis flare resolves. |
| | Fish oil, Hypertension (142, 60%) | √ | • Patient is currently taking fish oil 1000mg daily for hypertension prevention.<br>• Fish oil 2000mg daily for hypertension, follow-up outpatient blood pressure will be checked. | Niacin, Itching (92, 80%) | √ | • Allergen reactions, altace ramipril: nausea and diarrhea; niacin: itching and warm feeling.<br>• The patient has a strong family history of heart disease in his 30s and allergy to niacin with itching. |
| | Zinc, Wound | √ | • Would start 10 day courses zinc 220 mg/day for wound healing. | Niacin, Nausea (90, 70%) | √ | • Niacin causing nausea and decreased appetite. |
| | (71, 100%) | | For wound healing, recommend vitamin a 25,000 iu daily x 10 days, 50 mg zinc daily x 10 days. | | | Niacin – toxicity manifested primarily with ongoing epigastric discomfort, nausea and vomiting. |
| | Peppermint, Pain (48, 90%) | √ | She also has experienced pain relief when rubbing peppermint essential oil on the low back.<br>It was a slow process, but when she started on peppermint oils and water – it helped her pain better than zantac. | Fish oil, Nausea (57, 50%) | √ | Main reason for presenting today is to discuss an episode of epigastric pain and nausea, which occurred after ingesting a large dose of fish oil supplement.<br>He was supposed to take prescription strength fish oil capsules, but he took over the counter krill oil instead due to the nausea caused by the fish oil. |
| | Vitamin C, Anemia (36, 80%) | √ | Iron deficiency anemia – will continue with ferrous sulfate twice daily and instructed her to take with vitamin c to increase the absorption.<br>I will also ask her to be taking iron supplement and vitamin c to correct anemia as much as possible before surgery. | Fish oil, Diarrhea (35, 70%) | √ | Discussed titrating back up on fish oil as he tolerates, previously has been causing a lot of diarrhea so going slow.<br>Pt states he had diarrhea this fall, better since dc fish oil by pcp. |
| | Biotin, Hair loss (35, 100%) | √ | Patient is currently treating her hair loss with the following supplements: biotin 20,000 mcg daily and krill oil 1500 mg daily.<br>Patient has been taking biotin to control her hair loss. | Niacin, Flushing (21, 100%) | × | She was having significant flushing with niacin, so she discontinued this about 6 months ago.<br>We did discuss niacin as a potential strategy but I mentioned the difficulty with the flushing reaction. |
| | Vitamin E, Scar (19, 100%) | √ | Vitamin E po apply 1 capsule daily as needed to scar on forehead.<br>Apply Vitamin E to the scar for the next several months to help with it healing. | Niacin, Gi disturbance (20, 30%) | × | Allergen reactions: colesevelam hydrochloride: abdominal pain; niacin: gi disturbance, other see comments.<br>Allergen reactions: niacin: gi disturnbance; simvastation: cramps. |
| | Psyllium, Constipation (11, 100%) | √ | <ul><li>I suspect she has an underlying constipation which I recommend routine psyllium fiber starting at a low dose and titrating dose.</li><li>Patients states she takes psyllium powder daily for constipation, and needs refills.</li></ul> | Psyllium, Diarrhea (20, 40%) | √ | <ul><li>She was taking psyllium for constipation but had diarrhea.</li><li>His diarrhea may be related to the fact that he was on psyllium given his history of constipation.</li></ul> |

**Discussion**

Due to the inherent limitations of clinical trials and voluntary reporting data, the information regarding the DS safety and efficacy is incomplete and biased. With the increasing consumption and popularity of DS, there remains a critical need to expand our knowledge base of DS for patient safety, which is of extreme importance in the healthcare process. Clinical notes in EHR systems, documenting detailed and extensive real-world patients' information, present several advantages over conventional data sources, which can be leveraged for potential pharmacovigilance research. There are several studies demonstrating the utility of clinical notes in drug pharmacovigilance [33, 34], yet very few studies have attempted to investigate the use of clinical notes for monitoring the adverse events caused by DS. In this study, we have demonstrated the feasibility of automatic detection of DS safety signals in clinical notes using deep learning models. Without any external sources or feature engineering, our deep neural models have achieved better performance when compared with traditional machine learning models. Compared with studies investigating pharmacovigilance, our models also achieved comparable results [35].

When applied on the test dataset, the deep learning models demonstrate good generalizability. Using pretrained word embeddings as input, deep learning models can generalize well when used on unseen data because distributed word embeddings often carry semantic and syntactic relations between words. One of our previous studies [26] shows that word embeddings trained on a large medical corpus are capable of capturing the synonyms, brand names, abbreviations and misspellings of DS names. Using these distributed word embeddings as input, deep learning models can detect the named entities with similar word embeddings. However, for most clinical NLP systems, the NER component is mainly dictionary-based or traditional machine learning based. The dictionary-based NER system often has a high precision but low recall since the dictionary often fails to cover complete acronyms, abbreviation, and misspellings. It is well recognized that the performances of traditional machine learning models rely heavily on hand-crafted features. Determining the best set of features requires trial and error experiments. However, the deep learning models are totally end-to-end, with minimal work on feature engineering, which is more scalable and better for maintenance. Therefore, deep learning methods offer advantages over rule-based or traditional machine learning based methods. Additionally, the results also demonstrate that the combinations of word embeddings with character-level information is more informative than the word embeddings only. Interestingly, the large BERT model outperformed the Clinical BERT pre-training on MIMIC, and the potential reason may be that the MIMIC corpus (intensive care unit notes) does not sufficiently represent our corpus, collected from a variety of clinical settings. Moreover, Clinical BERT was trained from the BERT base model, which is smaller than the BERT large model.

During evaluation of generated pairs and source sentences in knowledge discovery, we found that the precisions for high frequency pairs are generally higher than those of less frequent pairs. Two types of errors were found during the error analysis. One type of error is that a sentence could contain several DS and AE, and the relation between one DS and one AE was wrongly predicted. For example, in the sentence "The patient is taking 1 tablet aspirin by mouth every 6 hours as needed for mild pain and fish oil considering the medical history of hypertension", the symptom "pain" was wrongly matched with the DS "fish oil". Another type of error is due to the preprocessing of clinical notes, which merges sentence from two sections into one sentence. For instance, in sentence "The patient is taking multivitamin po 1 tablet daily, fish oil 1000 mg po daily, past medical history: hemorrhage", the "past medical history" is the starting of another section in origin clinical notes. However, the fish oil is wrongly linked to the "hemorrhage" in the merged sentence.

The results of our study also show the feasibility of using clinical notes to perform real time DS safety monitoring. Applying the trained model on clinical notes can generate entity pairs of DS and AEs, which provide a new way for knowledge discovery or hypothesis generation. The valuable resources and knowledge obtained can help identify novel signals of AEs associated with DS. Such information can also assist subsequent in-depth investigations through clinical trials or in-vitro experiments by narrowing down the scope of DS, which can further optimize the use of DS and improve patients' safety.

There also exist some limitations of this study. The sample size for training the deep learning model is relatively small since manually annotating the clinical notes is expensive, labor extensive and time consuming. In the future, we may expand the data size and investigate how the increase of the data size will affect the deep learning model performance. Second, we did not consider the clinical terms that cross the sentence boundary. The third limitation is that we only included a small variety of features sets for training CRF and Random Forest. We may experiment with other syntactic, semantic, orthographic, and domain-specific features in the future work. For the NER deep learning models, we considered both word-level and character-level features. We may also train other traditional machine learning models such as SVM for performance comparison. However, one study [36] shows that the addition of word affixes information achieved better performance. Therefore, our future work might include affixes in our deep learning models.

## Conclusion

Automatic detection AEs related with DS use from clinical notes has a profound effect on patient safety. Deep learning models were applied to extract named entities of DS and events and their relationships from clinical notes in this study. When compared with traditional machine learning methods, the deep learning models have a better performance and generalizability. Our study has demonstrated that clinical notes hold great potential for monitoring the safety of DS use, which can create a new model for pharmacovigilance.

## Funding Statement

This work was supported by the NIH's National Center for Complementary & Integrative Health (NCCIH), the Office of Dietary Supplements (ODS) and National Institute on Aging (NIA) grant number R01AT009457 (PI: Zhang) and Clinical and Translational Science Award (CTSA) program grant number UL1TR002494 (PI: Blazar). The content is solely the responsibility of the authors and does not represent the official views of the NCCIH or ODS.

## Competing Interests Statement

The authors state that they have no competing interests to declare.

## Contributorship Statement

YF and RZ conceived the study idea and design. YF collected the data and performed the model training. SZ evaluated the results of knowledge discovery. YL and YF annotated the data. All author participated in writing and reviewed the manuscript. All authors read and approved the final manuscript.

## Acknowledgments:

We would like to thank Anusha Bompelli and Elizabeth Linderman for their help on the manuscript.

2. Oketch-Rabah HA, Roe AL, Muldoon-Jacobs K, Giancaspro GI. Challenges and Opportunities for Improving the Safety Assessment of Botanical Dietary Supplements: A United States Pharmacopeia Perspective. Clinical Pharmacology & Therapeutics. 2018 Sep;104(3):426-9.
3. Timbo BB, Chirtel SJ, Ihrie J, Oladipo T, Velez-Suarez L, Brewer V, Mozersky R. Dietary supplement adverse event report data from the FDA center for food safety and applied nutrition adverse event reporting system (CAERS), 2004-2013. Annals of Pharmacotherapy. 2018 May;52(5):431-8.
4. Harpaz R, DuMouchel W, LePendu P, Bauer-Mehren A, Ryan P, Shah NH. Performance of pharmacovigilance signal-detection algorithms for the FDA adverse event reporting system. Clinical Pharmacology & Therapeutics. 2013 Jun;93(6):539-46.
5. Poissant L, Taylor L, Huang A, Tamblyn R. Assessing the accuracy of an inter-institutional automated patient-specific health problem list. BMC medical informatics and decision making. 2010 Dec;10(1):10.
6. Uzuner Ö, Solti I, Cadag E. Extracting medication information from clinical text. Journal of the American Medical Informatics Association. 2010 Sep 1;17(5):514-8.
7. https://n2c2.dbmi.hms.harvard.edu. Retrieved on April 12 2020.
8. Jagannatha A, Liu F, Liu W, Yu H. Overview of the first natural language processing challenge for extracting medication, indication, and adverse drug events from electronic health record notes (MADE 1.0). Drug safety. 2019 Jan 21;42(1):99-111.
9. Grouin C, Deleger L, Zweigenbaum P. A simple rule-based medication extraction system. InProceedings of the Third i2b2 Workshop on Challenges in Natural Language Processing for Clinical Data 2009 Nov 13.
10. Roberts K, Rink B, Harabagiu S. Extraction of medical concepts, assertions, and relations from discharge summaries for the fourth i2b2/VA shared task. InProceedings of the 2010 i2b2/VA Workshop on Challenges in Natural Language Processing for Clinical Data. Boston, MA, USA: i2b2 2010.
11. Patrick J, Li M. A cascade approach to extract medication event (i2b2 challenge 2009). InProceedings of the Third i2b2 Workshop on Challenges in Natural Language Processing for Clinical Data 2009 Nov 13.
12. Chalapathy R, Borzeshi EZ, Piccardi M. Bidirectional LSTM-CRF for clinical concept extraction. arXiv preprint arXiv:1611.08373. 2016 Nov 25.
13. Habibi M, Weber L, Neves M, Wiegandt DL, Leser U. Deep learning with word embeddings improves biomedical named entity recognition. Bioinformatics. 2017 Jul 12;33(14):i37-48.
14. Bach N, Badaskar S. A review of relation extraction. Literature review for Language and Statistics II. 2007;2.
15. Dwyer JT, Coates PM, Smith MJ. Dietary supplements: regulatory challenges and research resources. Nutrients. 2018 Jan;10(1):41.
16. Rizvi RF, Vasilakes J, Adam TJ, Melton GB, Bishop JR, Bian J, Tao C, Zhang R. iDISK: the integrated DIetary Supplements Knowledge base. Journal of the American Medical Informatics Association. 2020 Apr;27(4):539-48.
17. Vasilakes J, Bompelli A, Bishop J, Adam T, Bodenreider O, Zhang R. Assessing the Enrichment of Dietary Supplement Coverage in the UMLS. Journal of American Medical Informatics Association. 2020 (accepted).
18. Fan Y, Pakhomov S, McEwan R, Zhao W, Lindemann E, Zhang R. Using word embeddings to expand terminology of dietary supplements on clinical notes. JAMIA open. 2019 Jul;2(2):246-53.
19. Vasilakes JA, Rizvi RF, Zhang J, Adam TJ, Zhang R. Detecting Signals of Dietary Supplement Adverse Events from the CFSAN Adverse Event Reporting System (CAERS). AMIA Summits on Translational Science Proceedings. 2019;2019:258.
20. Wang Y, Zhao Y, Bian J, Zhang R. Detecting Signals of Associations between Dietary Supplement Use and Mental Disorders from Twitter. In2018 IEEE International Conference on Healthcare Informatics Workshop (ICHI-W) 2018 Jun 4 (pp. 53-54). IEEE.
21. Wang Y, Gunashekar DR, Adam TJ, Zhang R. Mining adverse events of dietary supplements from product labels by topic modeling. Studies in health technology and informatics. 2017;245:614.
22. Vasilakes J, Fan Y, Rizvi R, Bompelli A, Bodenreider O, Zhang R. Normalizing Dietary Supplement Product Names Using the RxNorm Model. Studies in health technology and informatics. 2019 Aug 21;264:408.
23. Ryu JY, Kim HU, Lee SY. Deep learning improves prediction of drug–drug and drug–food interactions. Proceedings of the National Academy of Sciences. 2018 May 1;115(18):E4304-11.
24. https://github.com/nlpie/biomedicus3
25. Lample G, Ballesteros M, Subramanian S, Kawakami K, Dyer C. Neural architectures for named entity recognition. arXiv preprint arXiv:1603.01360. 2016 Mar 4.